\documentclass[prodmode,hillsideplop]{acmlarge}
\usepackage{flafter}
\usepackage[hidelinks]{hyperref}
\usepackage[hyphenbreaks]{breakurl}

\usepackage{svg}
\usepackage{float} 
\usepackage[width=.75\textwidth]{caption}

\acmVolume{22}
\acmYear{2022}
\acmMonth{10}

\usepackage[ruled]{algorithm2e}
\SetAlFnt{\algofont}
\SetAlCapFnt{\algofont}
\SetAlCapNameFnt{\algofont}
\SetAlCapHSkip{0pt}
\IncMargin{-\parindent}
\renewcommand{\algorithmcfname}{ALGORITHM}

\usepackage{todonotes}
\usepackage{xcolor}
\presetkeys%
    {todonotes}%
    {inline,backgroundcolor=yellow}{}

\title{Foundational DevOps Patterns}
\author{PAULO MARQUES \affil{Faculty of Engineering, University of Porto.}\ \\
FILIPE F. CORREIA \affil{Faculty of Engineering, University of Porto. INESCTEC.}}

\begin{abstract}
Adopting DevOps practices is nowadays a recurring task in the industry. DevOps is a set of practices intended to reduce the friction between the software development (Dev) and the IT
operations (Ops), resulting in higher quality software and a shorter development lifecycle. Even though many resources are talking about DevOps practices, they are often inconsistent with each other on the best DevOps practices. Furthermore, they lack the needed detail and structure for beginners to the DevOps field to quickly understand them.

In  order to tackle this issue, this paper proposes four foundational DevOps patterns: \textsc{Version Control Everything}, \textsc{Continuous Integration}, \textsc{Deployment Automation}, and \textsc{Monitoring}. The patterns are both detailed enough and structured to be easily reused by practitioners and flexible enough to accommodate different needs and quirks that might arise from their actual usage context. Furthermore, the patterns are tuned to the DevOps principle of Continuous Improvement by containing metrics so that practitioners can improve their pattern implementations.
\end{abstract}



\acmformat{P.\ Marques and F.\ Correia,  C. 2022. Foundational DevOps Patterns.}

\copyr{PLoP'22, OCTOBER 17--24, Virtual Online. Copyright 2022 is held by the author(s). {HILLSIDE} 978-1-941652-18-3}

\begin{document}

\begin{bottomstuff}
This work is supported by the Master in Software Informatics and Computation of the Faculty of Engineering of the University of Porto.\\
Authors' e-email addresses: P. Marques, paulodsam@gmail.com; F. F. Correia, filipe.correia@fe.up.pt.\\

Permission to make digital or hard copies of all or part of this work for personal or classroom use is granted without fee provided that copies are not made or distributed for profit or commercial advantage and that copies bear this notice and the full citation on the first page. To copy otherwise, to republish, to post on servers or to redistribute to lists, requires prior specific permission. A preliminary version of this paper was presented in a writers' workshop at the 27th Conference on Pattern Languages of Programs (PLoP).
\end{bottomstuff}

\maketitle

\section{Introduction}

The concept of DevOps was introduced in 2009~\cite{humble_why_2011} and quickly became popular, as it promises teams to deliver value faster while keeping the same quality level~\cite{accelerate_2018,StateOfDevOps}. In fact, adopting DevOps has been shown to lead to less burnout and an increase in job satisfaction~\cite{dora_webpage} and is now very common in the software industry.
The word DevOps comes from joining \textit{Development} with \textit{IT Operations}, and although DevOps has been discussed for many years now, there is still no academically accepted definition \cite{dyck_towards_2015}. This contrasts with the plethora of definitions that is possible can find online. One definition proposed by Leite~\textit{et al.} \cite{Survey_DevOps_Challenges} is:

\begin{quote}
    \textit{"DevOps is a collaborative and multidisciplinary effort within an organization to automate continuous delivery of new software versions while guaranteeing their correctness and reliability."}
\end{quote}

However, the lack of a consensual definition has led to an increasing diversity of software development practices being described under the general term of DevOps~\cite{lwakatare2016exploratory}. 
This, together with the fact that it often requires complex changes in organizations' processes and workflows~\cite{bucena_simplifying_nodate} makes its adoption anything but trivial.

Furthermore, a quick online search reveals several resources suggesting best practices for DevOps, but these practices are often not sufficiently explained or documented in such a way that they can be learned quickly and communicated effectively.

For these factors, organizations seeking to adopt DevOps often endure additional costs, as they feel the need to hire experts on the topic, or unnecessary costs such as paying to use the wrong tools for the wrong job due to misguidance.
Therefore, facilitating an efficient DevOps adoption by providing extra guidance on the adoption process becomes of the utmost interest.

A possible way to tackle this issue is by formalising the DevOps best practices as patterns. These provide sufficient detail so that people can more easily understand and apply the practices contained in them to the context of their organisation. With this goal in mind, we decided to gather a collection of patterns which we refer to as \textit{foundational}. By foundational, we mean a collection of patterns that are meant as a starting point for new DevOps adopters.
As such, we propose four new patterns that do not, by themselves, define DevOps but which provide a foundation for a broader pattern collection regarding DevOps practices and their adoption. In other words, they allow to generate the environment that enables such practices.

For mining and writing the patterns proposed in this paper we use three different sources of information for pattern mining:
\begin{itemize}
    \item \textbf{Literature} -- By performing the literature review in Section~\ref{chap:lr} we acquired different perspectives and definitions of the various DevOps practices/patterns. As such, these works provide the foundation for our pattern elaboration.
    \item \textbf{Grey literature} -- Many DevOps practices are shared through grey literature. They may appear in blog posts, DevOps tools' documentation, wikis made by companies and more. As such, they also provide valuable information for the writing of the patterns.
    \item \textbf{Past Experience} -- To add some details that the two previous kinds of literature do not possess, we use our experience using and implementing DevOps practices to fill in the blanks.
\end{itemize}

As alternative sources of information, we could have either mined patterns from existing tools or organized interviews with DevOps experts to check what their input was. These were mainly left out as a way to focus the scope of the work. Furthermore, we hypothesised that some of the information that we would get from both sources would already be partially present in the ones we used, mostly on grey literature. Usually the companies that provide the DevOps tools also have documentation and blog post regarding DevOps adoption. Furthermore, usually experts share their knowledge online, through blog post and others. As such, we felt that by not using these sources, we would not miss the bulk of the information, but it still might be worth while explore these sources in future research.

\section{Literature Review}\label{chap:lr}

We reviewed the literature with the goal of finding what practices it describes, and which of these have already been formalized in pattern format. We resorted to Google Scholar.
We also considered grey literature since it is common for practitioners and companies to write their knowledge on blogs or forums. In order to find it we used Google's search engine to find some of the results and from those we followed their references to other grey literature sources. To find works regarding patterns and practices for DevOps we used the following query:
\begin{quote}
\texttt {devOps AND (practices OR patterns OR capabilities)}
\end{quote}

After having the search results, we start by selecting the most relevant publications based on title, abstract, date, and methodology (if applicable).
Then we recorded the relevant ones in a \textit{to-read} list. Finally, we read and analysed the items on the list, looked up the references, and added the relevant ones to this list, including references to grey literature. If the work under analysis was part of grey literature, related works by the same author/organisation were also considered.

We started this analysis by understanding that a relevant number of patterns have been written to document practices for developing Cloud native systems~\cite{Richardson2018,RichardsonWebsite2021,Sousa2018c,Sousa2018b,Sousa2018a,Sousa2017,Sousa2015,sousa2016,Maia2022,Albuquerque2022}. Despite the proximity of these works to DevOps and to what it enables, we realised that most of these works do not specifically address DevOps practices and, therefore, we do not consider them in this analysis of related works. 

We found two works that tried to describe DevOps patterns. One of them is the website titled \textit{Cloud Adoption Patterns}, which provides a collection of patterns for the Cloud gathered from papers reviewed and accepted at xPLoP conferences. They present a collection of eleven patterns for developing, testing, and managing a cloud-native architecture. Furthermore, the authors present twelve more patterns around building a DevOps pipeline and managing containers~\cite{noauthor_patterns_nodate}. On the other hand, their focus is on DevOps patterns that support cloud adoption. As such, some DevOps patterns might not be regarded if they are not relevant for Cloud adoption. The other one is a collection of thirteen DevOps-related patterns, mined from Portuguese startup companies~\cite{teixeira_towards_2016}. However, we found that the context of some of these patterns lacks enough detail for an easy use.

We were able to find a variety of works regarding DevOps practices. The detail of the description for these practices ranges from a one-sentence description of the practice to somewhat detailed descriptions containing a few paragraphs as descriptions and tips on how to adopt it. However, there is an observable discrepancy in the extent of practices proposed by each author. This reflects the absence of consensus on the definition of DevOps or where the boundaries between other concepts should be set.
Table~\ref{tab:patternOrigin1} gathers all the practices that we have found, to help analyse them as a whole.

\begin{table}[H] 
    \centering
    \caption{DevOps practices and the corresponding works}
    \label{tab:patternOrigin1}
    \begin{tabular}{ l | l}
     Practice  & Works\\
     \hline
     Deployment automation & \cite{dora_webpage,lwakatare_fivecompany_caseStudy}\\
     Trunk-based development & \cite{dora_webpage,lwakatare_fivecompany_caseStudy}\\
     Shift left on security & \cite{dora_webpage}\\
     A loosely coupled architecture & \cite{dora_webpage}\\
     Empowering teams to choose tools & \cite{dora_webpage}\\
     Continuous integration & \cite{dora_webpage,pedra_devops_nodate,lwakatare_fivecompany_caseStudy}\\
     Continuous testing & \cite{dora_webpage}\\
     Version control & \cite{dora_webpage,pedra_devops_nodate}\\
     Test data management & \cite{dora_webpage}\\
     Comprehensive monitoring and observability & \cite{dora_webpage,lwakatare_fivecompany_caseStudy}\\
     Proactive notifications & \cite{dora_webpage}\\
     Database change management & \cite{dora_webpage}\\
     Code maintainability & \cite{dora_webpage}\\
     Continuous Delivery & \cite{dora_webpage,pedra_devops_nodate}\\
     Continuous Deployment & \cite{pedra_devops_nodate}\\
     Automated Build & \cite{pedra_devops_nodate}\\
     Cloud Infrastructure & \cite{dora_webpage,pedra_devops_nodate}\\
     Automated and Continuous Feedback & \cite{pedra_devops_nodate}\\
     Infrastructure as Code & \cite{pedra_devops_nodate,lwakatare_fivecompany_caseStudy}\\
     Change-based Code Review & \cite{lwakatare_fivecompany_caseStudy}\\
     Agile/Lean Practices & \cite{lwakatare_fivecompany_caseStudy}
    \end{tabular}
    
    \end{table}

\newpage
As we can observe, there are a few practices that have different names but appear to be similar or to overlap considerably, such as the case of \textit{Comprehensive monitoring and observability} \cite{dora_webpage} and \textit{Continuous monitoring}~\cite{pedra_devops_nodate} or \textit{Cloud infrastructure}~\cite{dora_webpage} and Cloud computing~\cite{pedra_devops_nodate}. Furthermore, some of the practices only appear in one work. 

We decided to refine this list to get a more consistent list of practices. First, we set out to find similar practices that, despite having different names, had a close-enough description that we could consider them the same practice. Afterwards, we decided to map the existing patterns to the practices to see if any of the practices appeared at least partially as part of a pattern or vice-versa. Then, we decided to exclude the practices that were only mentioned in one work and had no associated patterns. This was done so we could limit the scope of this work and focus only on the more consensual practices. 

Finally, we further analysed the remaining eight practices. One of them, \textit{Agile/Lean Practices}, was too broad for a pattern. As such, we removed it from the list. Then we have closely-related practices, such as \textit{Continuous Integration} and \textit{Trunk-based development} where the latter can be seen as a practice that enhances the first. While it is possible to do \textit{Continuous Integration} using other branching strategies, \textit{Trunk-based development} eases and allows to get the best of \textit{Continuous Integration}, as it supports faster feedback on the changes made to the code. Then we have the relation between \textit{Infrastructure as Code} and \textit{Version Control} where the first can be seen as a part of the latter. To address this, we decided to combine these practices into broader practices that would become our pattern candidates. In the end, we obtained five pattern candidates represented in Table~\ref{tab:patternOriginFinal}

\begin{table}[H] 
    \centering
    \caption{Pattern candidates, together with the works where we have identified the underlying practices (\textit{cf}. second column) and related DevOps patterns that we find in the literature (\textit{cf}. third and fourth columns).}
    \resizebox{\textwidth}{!}{%
    {\def\arraystretch{1.5}\tabcolsep=5pt
    \begin{tabular}{p{0.25\textwidth}|p{0.25\textwidth}|p{0.25\textwidth}|p{0.25\textwidth}}
     Practice & Works & Teixeira \textit{et al.} \cite{teixeira_towards_2016} & CAP \cite{noauthor_patterns_nodate}\\
     \hline
     Deployment automation & \cite{dora_webpage,lwakatare_fivecompany_caseStudy} & Deploying new instances & Automate VM Deployment\\
     Continuous integration & \cite{dora_webpage,pedra_devops_nodate,lwakatare_fivecompany_caseStudy}&Continuous Integration&\\
     Version control & \cite{dora_webpage,pedra_devops_nodate}&Version Control Organization, Reproducible Environments&\\
     Comprehensive monitoring and observability & \cite{dora_webpage,lwakatare_fivecompany_caseStudy}&Auditability, Alerting&Correlation ID\\
     Cloud Infrastructure & \cite{dora_webpage,pedra_devops_nodate}&Cloud, Scalling&Autoscaling, One Region, Overlay Network, Three Datacenters\\
    \end{tabular}%
    }
    }
    \label{tab:patternOriginFinal}
    \end{table}

\section{About the Patterns}\label{patterns:aboutPatterns}

\begin{figure}[H]
  \begin{center}
    \leavevmode
    \includegraphics[width=0.8\textwidth]{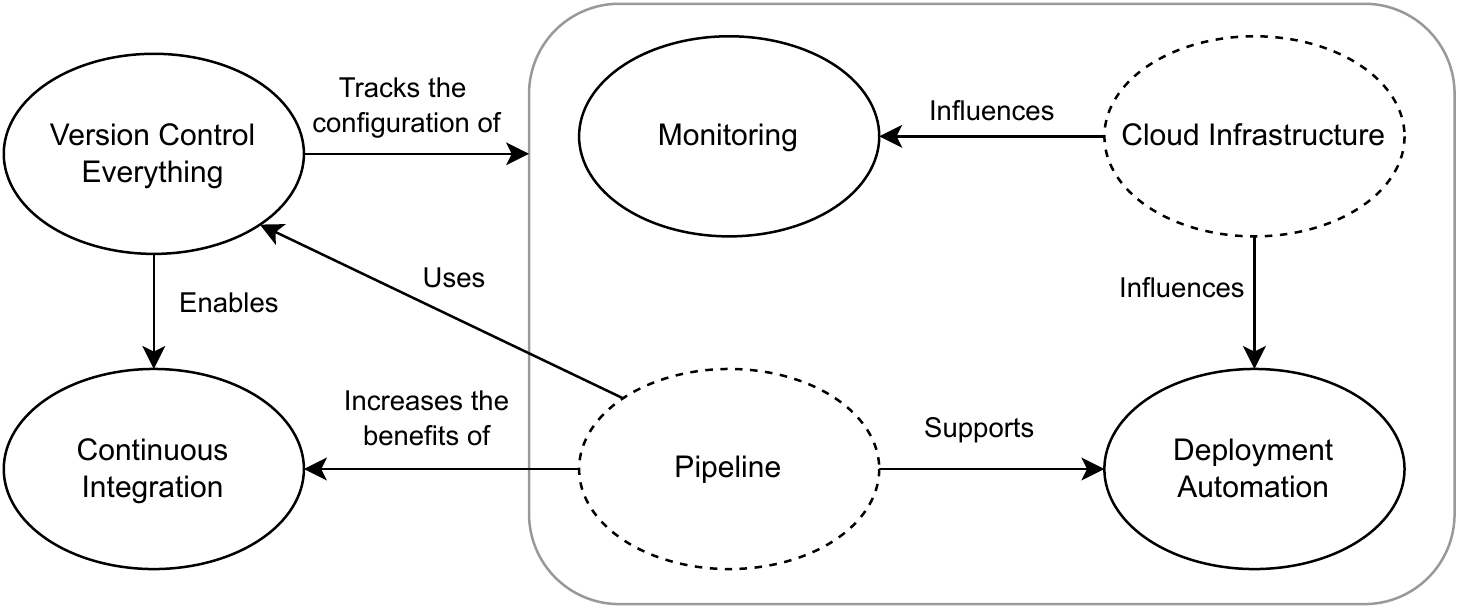}
    \caption{Overview of the pattern candidates and their relations}
    \label{fig:patterns}
  \end{center}
\end{figure}

We describe, in this paper, four of the five pattern candidates identified in the end of the previous section (\textit{cf}. Table~\ref{tab:patternOriginFinal}), deciding to leave out of the scope of this particular paper the \textsc{Cloud Infrastructure} pattern. The five pattern candidates and their relationships are portrayed in Figure~\ref{fig:patterns}.
Furthermore, we decided to include the \textsc{Pipeline} pattern in this figure, even though we had not derived it directly as an independent practices from the literature. However, we did find it mentioned in the literature as important for implementing \textit{Deployment Automation} and \textit{Continuous Integration}, and segregating it from the others allows us to make the solutions of these patterns easier to use, namely in contexts where a pipeline does not need to be present.
Finally, since the \textsc{Pipeline} pattern is already very well explained in other works \cite{Humble2010-mz,yoder2018delivering}, we decided it was not worth to describe it again here.

We opted to follow a structure that is very similar to the patterns in the website \textit{Cloud Adoption Patterns}~\cite{noauthor_patterns_nodate} with only small differences. Our patterns contain the following parts:

\begin{itemize}
	\item \textbf{Name:} An evocative name for the pattern.
	\item \textbf{Context:} Contains the context for the pattern providing a background for the problem.
	\item \textbf{Problem:} A question representing the problem that the pattern intends to solve.
	\item \textbf{Forces:} A list of forces that the solution must balance out.
	\item \textbf{Solution:} A detailed description of the solution for our pattern's problem.
    \item \textbf{Consequences:} The implications, advantages and trade-offs caused by using the pattern.
\end{itemize}

Similarly to the patterns by Brown et al.~\cite{noauthor_patterns_nodate}, we do not use explicit section headings to break these different parts but rely on some formatting conventions (\textit{e.g.} the solution is always described after a \textit{"therefore"}).
Our pattern structure diverges from the ones on the website because we decided to identify the following sections explicitly and display their content in bullet points:
\begin{itemize}
	\item \textbf{Related Patterns:} Patterns which are connected somehow to the one being described.
	\item \textbf{Metrics:} A set of metrics to measure the effectiveness of the pattern's solution implementation.
\end{itemize}

It is also worth noting that having a metrics section is not common in most patterns' structures. In fact, we could not find an equivalent section in the other works. However, we decided to add it since it supports a key principle of DevOps: continuous improvement.  While it is often hard to measure the \textit{consequences} of a pattern but they translate the actual benefits or liabilities of using that pattern, metrics are easy to measure but are a proxy to how well or thoroughly the pattern is implemented, or to which extent its consequences can be observed tangibly. They can be less meaningful but more actionable. A pattern's metrics should allow us to measure and plan improvements in the actual implementation of the pattern.

\pagebreak
\section{Version Control Everything}\label{patterns:vc}
Nowadays, developing and operating software is a multi-disciplinary job requiring people with different skills to work on different tasks assigned to them. To fulfil these tasks, they might need to create or modify different files such as configuration files, and source code among others. As such, they work on multiple artefacts that depend on each other. One example of this is the relationship between the development team and the operations team. Even though their jobs have different goals, they need to work on the same artefacts.

\textbf{How do you enable the simultaneous development and operation of multiple artefacts while keeping the consistency between them?}

Usually, a version of a system is composed of multiple artefacts such as application code, dependencies, configuration files or infrastructure-related scripts. They may be worked on independently, by different members of the team that is collaboratively working on the system, but this may make it harder to maintain consistency among the different artefacts. To ensure it, we may need to add extra steps and checks to the developing process, slowing it down.

Furthermore, teams often need to reproduce an environment in the instance of disaster recovery or trying to scale the software horizontally. As such, they need a way to keep track of the multiple artefacts and system versions to guarantee reproducibility. On the other hand, guaranteeing that the system versions are reproducible does not come effortlessly. You have to have the extra effort to consistently and correctly keep track of the system versions and which particular version of the artefact is present in that system version. You may manually keep the records, which might work on smaller and simpler systems, but easily becomes unfeasible in more complex systems with many artefacts.

Another factor that comes into play is that teams often have to demonstrate the system's integrity for auditability purposes. Therefore, teams need to have the ability to pick any release of the system and quickly determine the version of every artefact used to create it. Furthermore, they also need the ability to compare two releases of the system and determine what changed. As such, they need a versioning strategy that also guarantees traceability. However, these two common requirements when done manually are labour-intensive tasks.

Therefore,

\textbf{The use of a version control system for all production artefacts, including application code and its dependencies, application configurations, system configurations, scripts for automating the build and configuration of environments, and the configuration files for infrastructure specifications.}
In the version control system, teams must be able to query the current (and historical) state of their environments.

As a consequence, teams who employ this pattern have better disaster recovery capabilities when compared to teams who only version control some of the artefacts. Furthermore, it provides better auditability and traceability capabilities since teams must keep the history of all the artefacts on the version control system. This, also, enables better response to defects, since it provides the ability to roll back to a previous version when a critical defect or vulnerability is discovered.

As a downside, teams have to have the extra effort by defining, learning and using the version control strategy and maintaining a clear and consistent history of the artefacts. As such, they have to keep up-to-date documentation on how to use their version control system according to their strategy.

\subsection*{\textbf{Related Patterns}}
\begin{itemize}
\item \textsc{Continuous Integration}: Provides the inputs for the build and testing of the application. Furthermore, it facilitates the process of fixing a broken build by providing the ability to roll back to a previous version.
\item \textsc{Deployment Automation}: The scripts and the configuration information of the deployment process should be tracked in a version control system to ensure it is possible to recreate any environment in the event of disaster recovery.
\item \textsc{Cloud Infrastructure}: The cloud infrastructure configuration should be tracked in a version control system.
\item \textsc{Monitoring}: The monitoring configuration should be tracked in a version control system.
\item \textsc{Pipeline}: The implementations of the \textsc{Pipeline} pattern are strongly related to the version control system since the code that is supposed to run the pipeline is tracked in Version Control. As such, the pipeline implementation may depend on how you do Version Control. Additionally, the pipeline itself may be configured by artefacts tracked in the version control system.
\end{itemize}

\subsection*{\textbf{Metrics}}

The following metrics may be collected to measure the effectiveness of the implementation of this pattern:
\begin{itemize}
    \item The percentage of Application code in version control.
    \item The percentage of System configurations in version control.
    \item The percentage of Application configuration in version control.
    \item The percentage of Scripts for automating build and configuration in version control.
\end{itemize}

\section{Continuous Integration}\label{patterns:ci}

A team of developers is working on the same system or related systems. The system under development has a main version of the codebase (known as trunk, main, or mainline) and, as developers work on the system, diverging changes might appear. 

\textbf{How often should developers integrate their work with the mainline to optimize for feedback?}

Software systems are complex, and an apparently simple, self-contained change to a single file can have unintended side effects on the overall system. When a large number of developers work on the same system or related systems, coordinating code changes is a hard problem, as changes from different developers can be incompatible and it is difficult for team members to have visibility into what others are doing/changing. This results in problems from code conflicts to unexpected security concerns. Furthermore, the cost of merging diverging artefacts increases rapidly the more the artefacts to be merged differ from each other. 

However, if you try to integrate your changes as fast as possible, you might introduce regressions if you are not careful. There is, also, the added complexity of having to break down features into smaller chunks of code instead of delivering them all at once. Here a decision has to be made if there are enough changes to integrate and get feedback, or if the changes are too small to be meaningful.

Therefore,

\textbf{Developers do mainline integration as soon as they have a healthy commit they can share.}
As such, developers regularly integrate the code in short-lived branches or directly to the mainline/trunk. Ideally, a healthy commit equates to less than a day's work that has been tested and reviewed by a second person. To make the best of this, each integration triggers a build of the software and a  set of quick tests to discover regressions. In the event that the build fails or regressions are found, developers must fix them immediately. This can be easily achieved with the usage of a CI/CD \textsc{Pipeline}. The combination of these allows teams to integrate quicker, thus testing quicker resulting in a shorter feedback loop, thus making software delivery faster. 

As a consequence, teams who adopt this pattern will integrate more often, which means they can test more often, resulting in quicker feedback loops. Testing often results in less likelihood of regressions occurring during development, and, thus, higher quality software. Furthermore, integrating often leads to team members having more visibility into what others are working on, reducing even the likelihood of regressions from diverging changes.

Also, integrating more often and with smaller chunks of code makes it simpler to solve the merges of diverging changes. This allows teams to allocate more time to the development of new features and, thus, making them more productive. 

On the other hand, teams are obliged to maintain not only the code but also a test suite resulting in added complexity in the development process. 

Furthermore, there is additional complexity in the development process by having to break down features into smaller incremental steps.

\subsection*{\textbf{Related Patterns}}
\begin{itemize}
\item \textsc{Version Control Everything:} The pattern presented here relies on the existence of a version control system as well as the ability to quickly fix broken builds by rolling back to a previous version.
\item \textsc{Pipeline:} Even though you technically do not need a pipeline to adopt \textsc{Continuous Integration}, pipelines potentiate the benefits of continuous integration, as they enable richer feedback about the changes you integrate in the code.
\end{itemize}

\subsection*{\textbf{Metrics}}

The following metrics may be collected to measure the effectiveness of the implementation of this pattern:
\begin{itemize}
    \item The percentage of code commits that result in a software build.
    \item The percentage of commits that are automatically tested.
    \item The percentage of successfully executed automated builds every day.
    \item The percentage of successfully executed automated tests every day.
    \item The percentage of builds available for testing.
    \item The percentage of acceptance tests' results that are available within a day.
    \item The time it takes between the build breaking and having it fixed.
\end{itemize}

\section{Deployment Automation}\label{patterns:DA}

The development team has decided to deploy the software in a new environment, such as testing or production. As such they reach out to the operations team as this is a combined effort of the two. There is already infrastructure to support the environment. 

\textbf{How to deploy software in a reproducible, consistent and timely way?}

Depending on the target environment and the type of software, the deployment process' complexity might range from simple like executing only a program to a very complex process containing various steps.

Manually deploying a system is an easier task when compared to automating its deployment. Depending on the system, it might be really time-consuming to automate its deployment when compared to manual deployment. Adding to this fact, deployments might be rare events for certain organizations. However, if the organization is trying to accelerate the speed with which they deliver features, they should expect to deploy more often, and, as such, deployments become more frequent.

Furthermore, when manually deploying, there is a higher risk of mistakes occurring during the deployment process, such as misconfigurations. This may lead to failed deploys when the new version of the software being deployed did not cause the issues. Therefore, debugging effort must be put into the failed deployment, which presents an obstacle to faster feedback. As such, the deployment process should be reliable. On the flip side, if your system is simple enough, the chance that manually deploying introduces problems might be low.

Finally, depending on the deployment complexity, different people deploying the software might do things slightly differently, obtaining different results and making deployments not reproducible. As a consequence, if the need comes to redeploy in the event of a disaster recovery scenario or even for manual testing and validation of the software, recreating the state of the environment might be extremely hard. However, if the system is simple enough, recreating the state might also be trivial.

Therefore,

\textbf{The deployment process should be automated as much as possible.}

The first step is to document the typical deployment process in a tool that both the development and operations have access to. From there, work on simplifying and automating the deployment process, implementing idempotence and order independence wherever possible. If available, the capabilities of the infrastructure platform should be leveraged. The goal is to make deployments as straightforward as possible.

As a consequence, having an automated process to deploy is usually faster than a manual one. As such, teams do not have to dedicate time to the deployment part, they can allocate it to other more valuable tasks, thus, increasing the productivity of the teams. Furthermore, teams no longer have to wait for someone to start manually deploying to find issues in the software changes related to the deployment process, shortening the feedback loop and allowing teams to fix these defects more quickly.
On the other hand, automating the deployment process is not a trivial task. Depending on the complexity of the software, automating every step of the process might prove complex and time-consuming.

\subsection*{\textbf{Related Patterns}}

\begin{itemize}
\item \textsc{Version Control Everything}: The scripts and the configuration information of the deployment process should be tracked in a version control system to ensure it is possible to recreate any environment in the event of disaster recovery.
\item \textsc{Cloud infrastructure:} Cloud infrastructure often simplifies the deployment process by having built-in tools for that purpose.
\item \textsc{Pipeline:} It is possible to have \textsc{Deployment Automation} without \textsc{Pipeline} but the latter supports and makes it easier to do the former.
\end{itemize}

\subsection*{\textbf{Metrics}}

The following metrics may be collected to measure the effectiveness of the implementation of this pattern:
\begin{itemize}
    \item Number of manual steps in the deployment process.
    \item Time spent on deployment pipeline's delays.
\end{itemize}


\section{Monitoring}\label{patterns:monitoring}

A group of developers is planning the deployment or has already deployed the software system they are working on into the production environment.

\textbf{How do you allow teams to understand and actively diagnose the health of their systems?}

When working with complex systems that are already in production, there is the risk that changes or added functionality may introduce unexpected side effects that may have escaped the tests. However, changes are usually welcomed since they usually provide extra value to the clients.

In the instance of an outage or service degradation teams need to find what is causing the issues and how to fix them promptly. In other words, teams need to minimize the time to restore (TTR). However, teams want to avoid the need to restore a system altogether. As such, teams must detect possible service degradation and outages before they occur. However, these might prove a time-consuming task, requiring the allocation of multiple resources.

On the other hand, enabling the previously mentioned capabilities increases the development and operational complexity of the system. It is easier to design and develop a system without having observability in mind. However, it is harder to achieve the above capabilities if the system was not designed having observability in mind. Reworking the system to make it more observable might prove difficult and costly.
Therefore,\\
\textbf{Implement a comprehensive monitoring system that allows teams to monitor predefined metrics on the overall health of the system and the system state as experienced by the users. Furthermore, it should provide the tooling that enables teams to actively debug the system and make it observable.}
When the system detects an issue or predicts that one is about to arise, it should preemptively alert the teams of the issue. You should monitor things like the state of the networks, the state of the infrastructure, the performance of systems, and key business and system metrics, among others.

As a consequence, it is possible to understand how the system is doing in production, getting faster feedback from the production environment, when compared to an approach where you would wait got problems to be reported by the end user. This allows the team to address problems before they impact the users resulting in a better quality service. Furthermore, with monitoring, it is possible to detect problems that would result in system downtime if not timely addressed.  

\subsection*{\textbf{Related Patterns}}
\begin{itemize}
    \item \textsc{Version Control Everything} Relies on the version control system to track the monitoring configuration.
    \item \textsc{Cloud Infrastructure} The monitoring strategy implemented is dictated by the use or not of cloud infrastructure
    \item The patterns by Albuquerque et al.~\cite{Albuquerque2022} provide further information on how to adopt a monitoring strategy.
\end{itemize}

\subsection*{\textbf{Metrics}}

The following metrics may be collected to measure the effectiveness of the implementation of this pattern over some time:
\begin{itemize}
    \item Number of alerts resulted in no action, or were marked as "Working as Intended"? (False positives).
    \item Number of system failures happened with no alerting, or alerting later than expected? (False Negative).
    \item Mean time to detect.
    \item Mean time to repair.
\end{itemize}

\section{Conclusions}
This paper proposes four new patterns for DevOps: \textsc{Version Control Everything}, \textsc{Continuouns Integration}, \textsc{Deployment Automation} and \textsc{Monitoring}. These patterns provided a guided method to facilitate the DevOps adoption while keeping some of the DevOps core principles, such as Continuous Improvement.

The proposed patterns were mined from existing literature, complemented grey literature, and with the authors' experience. Since the patterns are newly written, their description may not be complete or miss crucial details. In future work, we may further evaluate their accuracy, corroborating or refuting their existence, following some of our previous  research~\cite{Sousa2022,Vale2022}.
Finally, we do not believe that we suggest all the DevOps patterns and as such future work may consist of mining the remaining patterns for a complete pattern catalogue. Contrasting to the mining process we used here, future mining should look for patterns in the tooling used in the industry.

\section{Acknowledgments}

We would like to thank Alfredo Goldman, but also Matheus Bernardino and Leonardo Leite, for their comments and suggestions for improvement during the shepherding of this paper for PLoP 2022. We also thank Jessica Diaz, for her valuable feedback on early versions of this work. Finally, we would like to thank all the participants of our writers' workshop at PLoP 2022, who through their comments and discussion helped us to improve these patterns considerably---Dominique Causse, Eduardo Fernandez, Joelma Choma, Joseph Yoder, Marden Neubert and Pavel Hruby.

This work is financed by European Commission funds within the project "EuroCC H2020-JTI-EuroHPC-04-2019: HPC Competences Centres-Research and Innovation", through grant agreement 951732.

\bibliographystyle{ACM-Reference-Format-Journals}
\bibliography{references}

\end{document}